\documentclass[preprint]{aastex}

\def\<#1>{\langle\hbox{#1}\rangle}

\def\etal{{et~al.}}
\def\kms{km~s$^{-1}$}

\def\eg{{e.g.}}
\def\ie{{i.e.}}
\def\simgt{\hbox{\rlap{\raise 0.425ex\hbox{$>$}}\lower 0.65ex\hbox{$\sim$}}}
\def\simlt{\hbox{\rlap{\raise 0.425ex\hbox{$<$}}\lower 0.65ex\hbox{$\sim$}}}

\def\zl{z_{\rm L}} \def\zs{z_{\rm S}}

\begin{document}

\title{Is B1422+231 a Golden Lens?}

       \author{Somak Raychaudhury}
       \affil{School of Physics and Astronomy\\
              University of Birmingham\\
              Edgbaston, Birmingham B15~2TT, UK}
       \email{somak@star.sr.bham.ac.uk}
       
       \author{Prasenjit Saha}
       \affil{Astronomy Unit\\
              Queen Mary and Westfield College\\
              University of London\\
              London E1~4NS, UK}
       \email{p.saha@qmul.ac.uk}

       \and

       \author{Liliya L.R. Williams}
       \affil{Department of Astronomy\\
              University of Minnesota\\
	      116 Church Street SE\\
              Minneapolis, MN 55455}
       \email{llrw@astro.umn.edu}

\begin{abstract}
B1422+231 is a quadruply-imaged QSO with an exceptionally large
lensing contribution from group galaxies other than main lensing
galaxy.  We detect diffuse X-rays from the galaxy group in archival
Chandra observations; the inferred temperature is consistent with the
published velocity dispersion.  We then explore the range of possible
mass maps that would be consistent with the observed image positions,
radio fluxes, and ellipticities.  Under plausible but not very
restrictive assumptions about the lensing galaxy, predicted time
delays involving the faint fourth image are fairly well constrained
around $7h^{-1}\,\rm days$.
\end{abstract}


\section{Introduction}

One of the most attractive things about lensed quasars is the
possibility of measuring the distance scale without any local
calibrators.  If the source is variable and the variation is observed
in different images with time delays then a formula of the type
\begin{eqnarray}
\<time delay> &=& h^{-1} \zl(1+\zl) \nonumber \\
&\times& \<1 month> \times \<image separation in arcsec>^2 \\
&\times& \<lens-profile dependent factor> \nonumber
\end{eqnarray}
(where $\zl$ is the lens redshift) applies.  This formula depends
weakly on the source redshift (the time delay gets somewhat shorter if
the source is not very much further than the lens) and on cosmology
(for the same $h$ an Einstein-de Sitter cosmology gives long time
delays, an open universe gives short time delays, with the currently
favored flat $\Lambda$-cosmology being intermediate); but these
dependencies are at the 10\% level or less.  The troublesome term is
the lens-profile dependent factor, which is of order unity but for a
given lens can be uncertain by a factor of two or more.  So while the
basic theory is well established \citep{r64}, measuring $h$ from
lensing is still problematic.  Meanwhile the search for
well-constrained lenses continues, and lensed-quasar aficionados speak
of them as ``golden lenses'' \citep{gl97}.

In this paper, we study a particularly interesting lens: B1422+231 was
discovered in the JVAS survey \citep{discovery} and consists of four
images of a quasar at $\zs\!=\!3.62$ lensed by an elliptical galaxy at
$\zl\!=\!0.334$ \citep{hst96,castles} with several nearby galaxies at
the same redshift \citep{kun97,tonry98}.  It is especially interesting
for four reasons.  First, while external shear from other group
galaxies is important in most four-image lenses, unlike other
well-studied quads, B1422 is {\em dominated\/} by external shear.
Second, the lensing-galaxy group can be detected directly from its
X-ray emission.  Third, VLBI imaging of the quasar core \citep{pat99}
gives information on not just the flux ratios but on the tensor
magnification ratios.  And fourth, a time delay between two images has
been reported \citep{pat01}.

\begin{table}[b!]
\hrule\bigskip
\begin{center}
\caption{Summary of data used for lens reconstruction (The image
numbers are as shown in Fig.~1, the origin of
coordinates being ``G'').}
\begin{tabular}{crrcrr}
\noalign{\null\bigskip}
\tableline\tableline
\noalign{\smallskip}
image & \multicolumn1c{$x$} & \multicolumn1c{$y$}
& axis ratio  & \multicolumn1c{PA}
& \multicolumn1c{mag} \\
\noalign{\smallskip}
\tableline
\noalign{\smallskip}
1  & $  1.014 $ & $ -0.168 $ & 3--4  & $  16^\circ $ & $ 15 $ \\
2  & $  0.291 $ & $  0.900 $ & 5--7  & $  53^\circ $ & $ 30 $ \\
3  & $  0.680 $ & $  0.580 $ & 8--10 & $  43^\circ $ & $-30 $ \\
4  & $ -0.271 $ & $ -0.222 $ & 1--2  & $ 123^\circ $ & $ -1 $ \\
\noalign{\smallskip}
\tableline\tableline
\end{tabular}
\label{tbl}
\end{center}
\end{table}

\section{Summary of previous observations}

\subsection{Image configuration}

\begin{figure}[tbh!]
\epsscale{0.5}
\plotone{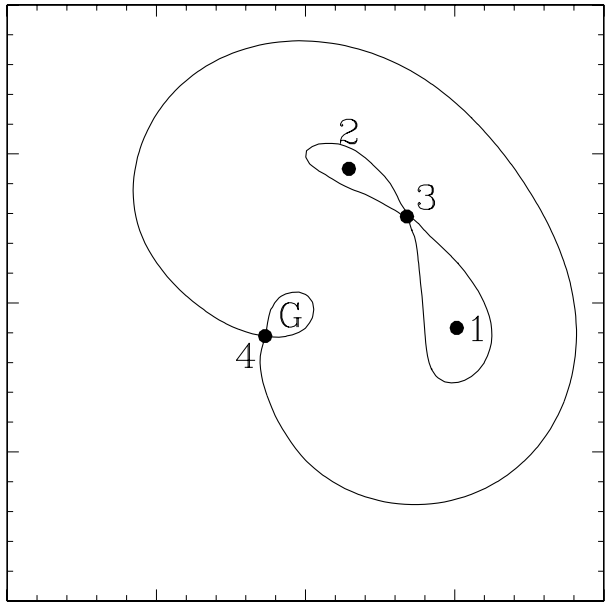}
\caption
{\footnotesize
Image configuration in B1422 (North is up, East is to the left).  
The filled circles labeled 1--4 mark
the images, ordered by arrival time; 1 and 2 are minima, while 3 and 4
are saddle points.  An unobservable fifth image would be a maximum at
the center of the lensing galaxy (near 'G').  The curves are
saddle-point contours in the arrival-time surface---see \citet{bn86}
for the significance of these. The precise location of the
saddle-point contours in this figure are model-dependent, but the
qualitative features are model-independent and generic for quads---see
\eg, Figure~1 of \citet{sw01}.
\label{schem}}
\bigskip\hrule
\end{figure}

The quad has three images with $V=16.5$ to $17$ nearly in a straight
line, and a fourth image with $V=20$ on the other side of the galaxy
(see Fig.~\ref{schem}).  The maximum image separation is $1.3''$.
Examining the image configuration and following the ideas of
\citet{bn86} on classification of images, we can work out some basic
properties of the system.

First, although time delays are very uncertain, and predictions are
model-dependent, the {\it ordering\/} of arrival times is easily
inferable from the image positions and is model-independent.  Thus the
images can be labeled 1,2,3,4, by increasing light-travel time.
Figure~\ref{schem} does so.  We will refer to individual images by
these time-order labels.

Second, the image configuration is noticeably elongated along
NE/SW. Such elongation generically indicates \citep{ws00} that the
lensing potential is significantly elongated along NW/SE.  Since the main
lensing galaxy is an elliptical, and cannot generate so much shear by itself,
we must immediately suspect a large external shear.

\subsection{External shear}\label{ext_shear}

Apart from qualitative evidence from the morphology, the importance of
external shear in this case is further supported by the observations
of galaxy group members, and a comparison with the PG1115 group. The
two groups are rather similar (see Kundi\'c et al.\ 1997 for
comparison) including number of galaxies in the group, and the
distance of the main lensing galaxy from the group's center, about
$15''$.  But the apparent magnitude, and presumably the associated
mass of $G$ in B1422 compared to magnitudes (and masses) of most of
the other group members is small ($V_G=21.5$ vs. $V_{G2}=20.4$,
$V_{G3}=20.0$, $V_{G4}=21.6$, $V_{G5}=20.3$), whereas in PG1115, $G$
is comparable in brightness to other group members ($R_G=20.2$
vs. $R_{G1}=19.0$, $R_{G2}=20.0$, $R_{G3}=20.5$). We conclude that
external shear is more important in B1422 than in PG1115, and most
other 4-image cases.

We can roughly estimate the magnitude of the shear.  For a group or a
cluster, the angular radius of the Einstein ring is
\begin{displaymath}
\theta_{\rm E} \sim 2\arcsec \times 
\frac{\langle \sigma_v^2 \rangle}{\left(300\ {\rm km\;s}^{-1}\right)^2},
\end{displaymath}
where $\sigma_v^2$ is the line-of-sight velocity dispersion of the
cluster. The external shear due to this cluster (modeled as an
isothermal sphere) is $\gamma\sim \theta_E/\theta$, where $\theta$ is
the distance of the group center from the main lensing galaxy. Depending 
on whether $G3$ is included or excluded in the determination of the
group's velocity dispersion, $\sigma_v$ is $550\pm 50\,$\kms~ or $240\pm
60\,$\kms.  Folding in the uncertainty in the location of the group's
center, between $10''$ and $14''$, the maximum and minimum estimates
of the external shear are $\gamma=0.67$ and $0.09$. Consistent with
these bounds, other workers have fitted models that require
$\gamma=0.25$ \citep{hb94} and $\gamma\sim 0.1$ \citep{ksb94}.  Using
an elliptical potential model \citet{wm97} derive a lower limit to the
external shear of 0.11. Thus, by all accounts, the external shear
plays a central role in B1422.

\subsection{Magnification}

The radio fluxes of the images 1,2,3 and 4 are in the ratio 16:30:33:1
at 8.4 GHz and similar at other radio frequencies \citep{pat99,ros01},
whereas the flux ratios in the optical and near-infrared are 8:13:16:1
\citep{yb96,ir92}. Reddening by the lensing galaxy is unlikely because
the flux ratios do not change much between different optical
wavebands.  Microlensing due to stars is also unlikely because the
flux ratio of images 2:3 stayed the same while both images changed in
brightness (Srianand \& Narasimha, private communication). This
suggests that millilensing may be taking place, \ie, lensing by
substructure of scale intermediate between individual stars and the
whole galaxy \citep{ms98}. If that is the case then radio fluxes are
probing larger structure in the galaxy than the immediate neighborhood
of the images.

Previous models,\footnote{Some models assume $\zl\!=\!0.6$ (based on
an earlier, incorrectly measured redshift) rather than the value we
use here, namely $\zl\!=\!0.334$ \citep{hst96}. The correction of the
redshift implies only a scale-change for these models. In particular,
the time delays predicted by these models must be multiplied by 0.52.}
while agreeing about the direction of external shear, have been unable
to to reproduce the observed flux ratios
\citep{hb94,ksb94,kks97,wm97}.

VLBI maps by \cite{pat99} show different shapes and sizes for each
image of the quasar core. It is tempting to identify the ellipticity
and area of the core in each image with the relative tensor
magnification; but the quoted area ratios of 1.34:1.47:1.34:1 are
very different from the flux ratios, so such an identification would
be incorrect.  Evidently, the mapped core in image 4 corresponds to a
larger area on the source than the mapped cores in images 1--3.
Still, we would like to incorporate some information from these
remarkable VLBI maps into our mass models.  So in this paper we will
adopt a compromise: we take the ellipticities of the mapped cores as
the ellipticity corresponding to the magnification tensor, but we
disregard the areas of the mapped cores and take the fluxes as
corresponding to the scalar magnification.  (How this works
operationally is explained below, with equation~\ref{tensormag}.)  The
image and magnification data we use for modeling are collated in
Table~\ref{tbl}.

We remark that since the unlensed flux is unknown, the absolute
magnifications are also unknown.  This unknown factor is a
manifestation of the mass disk degeneracy---see \citet{s00} and
references therein.

\section{Chandra X-ray observation of the group}

Redshifts of six galaxies, including that of the lens, indicate that
the lens is a member of a compact group whose line-of-sight 
velocity dispersion is 550 km/s and median projected radius 
$35\,h^{-1}$ kpc \citep{kun97}. 
A Faraday rotation corresponding to a rotation measure of
$280\pm 20$ rad~m$^{-2}$ \citep{pat99} between images 2 and 3 is much
larger than expected from an elliptical galaxy and further supports
the presence of a group around the lens.

\begin{figure}[tbh!]
\epsscale{0.8}
\plotone{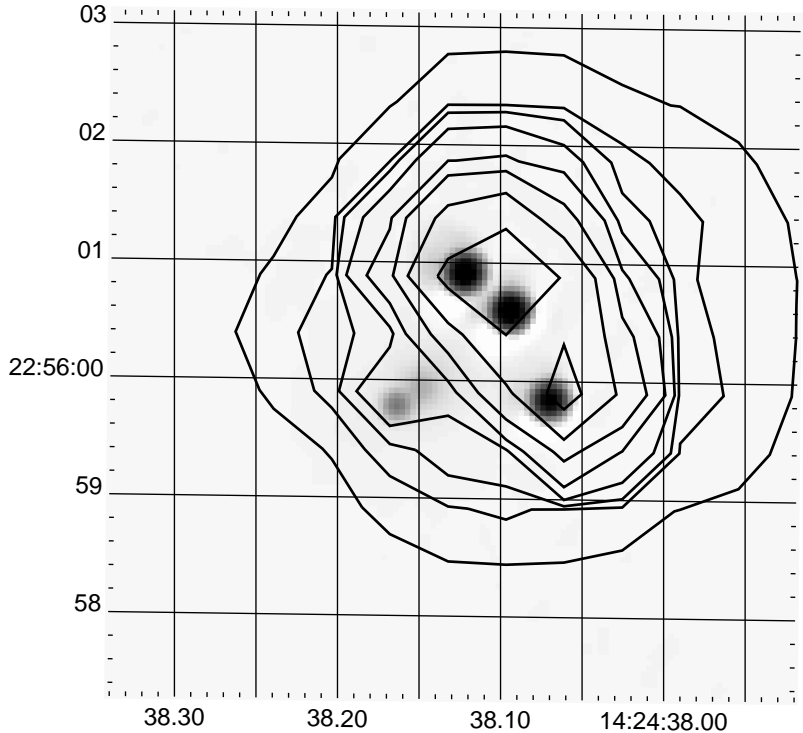}
\caption
{\footnotesize
Chandra ACIS-S observation of B1422+231. The X-ray observation is
represented as logarithmic contours superposed on the HST H-band
NICMOS image, kindly supplied by the CASTLES team.  Images 2-3 and 1
are separately detected, and there is a hint of image~4 being detected
as well. The pixel size of ACIS-S (0.49 arcsec) is not enough to
resolve images 2 and 3.  There is a slight astrometric offset between
the HST and Chandra images, which we have made no attempt to correct
for.
\label{fig:xray}}
\bigskip\hrule
\end{figure}

We show here the unambiguous detection of diffuse X-ray emission from
hot gas belonging to this group in a {\em Chandra}
observation. Previously, from a ROSAT HRI observation, \cite{siebri98}
had found ``no evidence of extended emission'' over and above the
$5^{\prime\prime}$ PSF of the HRI. Figure~\ref{fig:xray} shows the
28~ks Chandra ACIS-S observation represented as contours superposed on
the HST image taken with the 
NICMOS (H-band) by the CASTLES team \citep{fpcast}, where the four images,
and well as the lens galaxy, are distinctly seen. There is clearly an
astrometric offset between the HST and Chandra images, which we have
made no attempt to correct for. However, we can safely assume that the
two peaks in the X-ray distribution correspond to images 2 and 3
(brighter) and 1. The diffuse emission outside the source images is
softer, and extends to a scale of about 3~arcsec, which corresponds to
about $9\,h^{-1}$ kpc at the lens redshift $z=0.334$.

B1422 was observed with the ACIS-S instrument on board the {\it Chandra}
observatory on 2000 June 1 for 28.8 ks, the data being collected on
the back-illuminated S3 CCD. We applied the standard data
processing techniques recommended by the Chandra X-Ray Center (CXC), 
including the recent task ACISABS, to correct  for the absorption, 
predominantly at softer energies, caused by 
molecular contamination
of the ACIS optical blocking filters. Since the exposure time 
is long enough, we reprocessed the data without randomizing 
the event positions, which improved the point-spread function.
Periods of high background were filtered out, leaving 28.4 ks of
usable data.

\begin{figure}[tbh!]
\epsscale{0.7}
\plotone{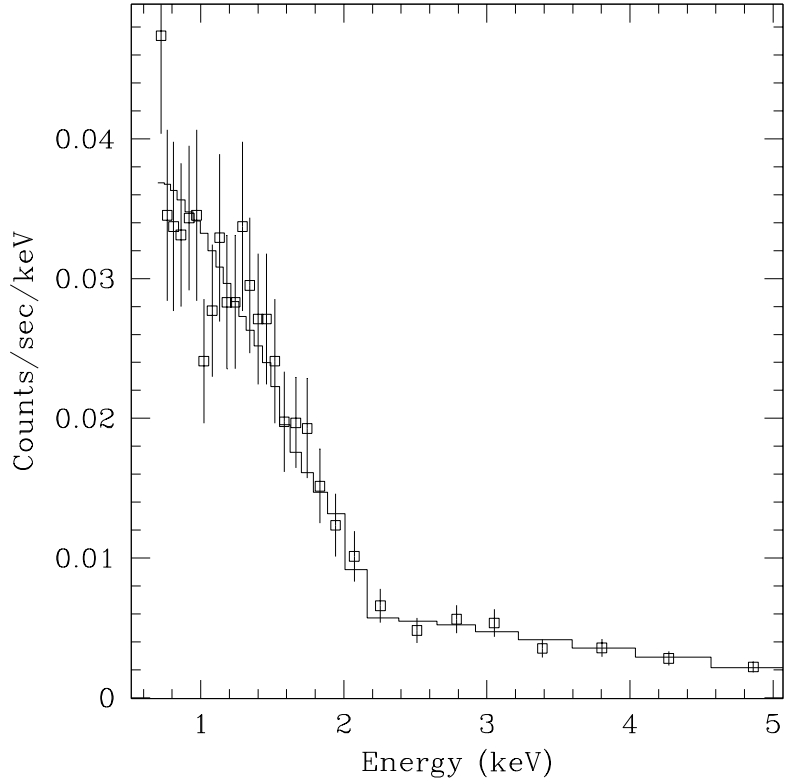}
\caption
{\footnotesize
Chandra ACIS-S spectrum of B1422+231 of the diffuse emission from 
the lens, excluding photons from regions of radius 0.7 arcsec around 
each of the three QSO image regions. The best-fit model to the data 
in the range 0.5-6 keV (solid histogram) consists of a Raymond-Smith 
thermal bremsstrahlung plasma, and a power-law for the source quasar, 
keeping the normalization of both components free.
\label{fig:xray-spec}}
\bigskip\hrule
\end{figure}

To separate the spectrum of the source quasar from that of the lensing
galaxy group, we isolated circular areas of radius 0.7 arcsec around
each of the three QSO images and extracted the spectrum of the
quasar. The power-law fit (together with Galactic absorption) to it
yielded a power-law of slope 1.41 with a reduced $\chi^2$ statistic of
0.9. There was no evidence of extra absorption at the redshift of the
quasar.  We then fixed the slope of the power law, and fitted a model
of thermal bremsstrahlung plasma (Raymond-Smith) plus the quasar
power-law spectrum to the rest of the X-ray emission, keeping the
normalization of both components free, but fixing elemental abundance
to 0.3 solar (reasonable for a group or cluster of galaxies). A fit to
the data in the range 0.5-6 keV yielded a temperature of 0.71 keV for
the group of galaxies, which translates into a velocity dispersion
range of 150-450 \kms~ \citep{hels00}, consistent with the observed
dispersion of 550, or 240\kms (depending on whether $G3$ is included
or not).  The fits were performed with the CXC software package
SHERPA.

The $M-T_X$ relation (mass vs. x-ray temperature)
for virialized isothermal systems of galaxies, for
mass within a radius $0.3\,R_{\rm virial}$, gives a mass of  $3.6\times
10^{12}\,M_{\odot}$ for the above temperature \citep{sand03}.  The
V-magnitude of the main lensing galaxy is $m_V=21.5$ \citep{kun97}, 
which after correction for colour evolution \citep{bru01} and 
$k$-correction \citep{fg94} yields a luminosity of 
$L_B=1.9\times 10^{10}\,h_{75}^{-2}\,L_{B,\odot}$ and a mass-to-light 
ratio of $M/L_B=190\,h_{75}$, far larger than that of a single galaxy. 
The bolometric X-ray flux of the diffuse emission is $L_X=2\times 10^{42}$
erg~s$^{-1}$, which means $\log (L_X/L_B)=32$, also much larger than
typical values for galaxies \citep{hels01}. We therefore conclude
that the diffuse emission detected by Chandra belongs to the group
of galaxies the lens is a part of.

\section{A simplified model}\label{simple}

Significant external shear combined with the fortuitous positioning of
the QSO source along the same line as the group and the center of the
main lensing galaxy (this alignment is a robust feature in all models 
for this system) transforms the case of images 3 and 4 of B1422 into a
one-dimensional problem. As we now show this makes the problem much
better constrained than a generic 4(+1) image system.

\begin{figure}[p!]
\epsscale{1.0}
\plotone{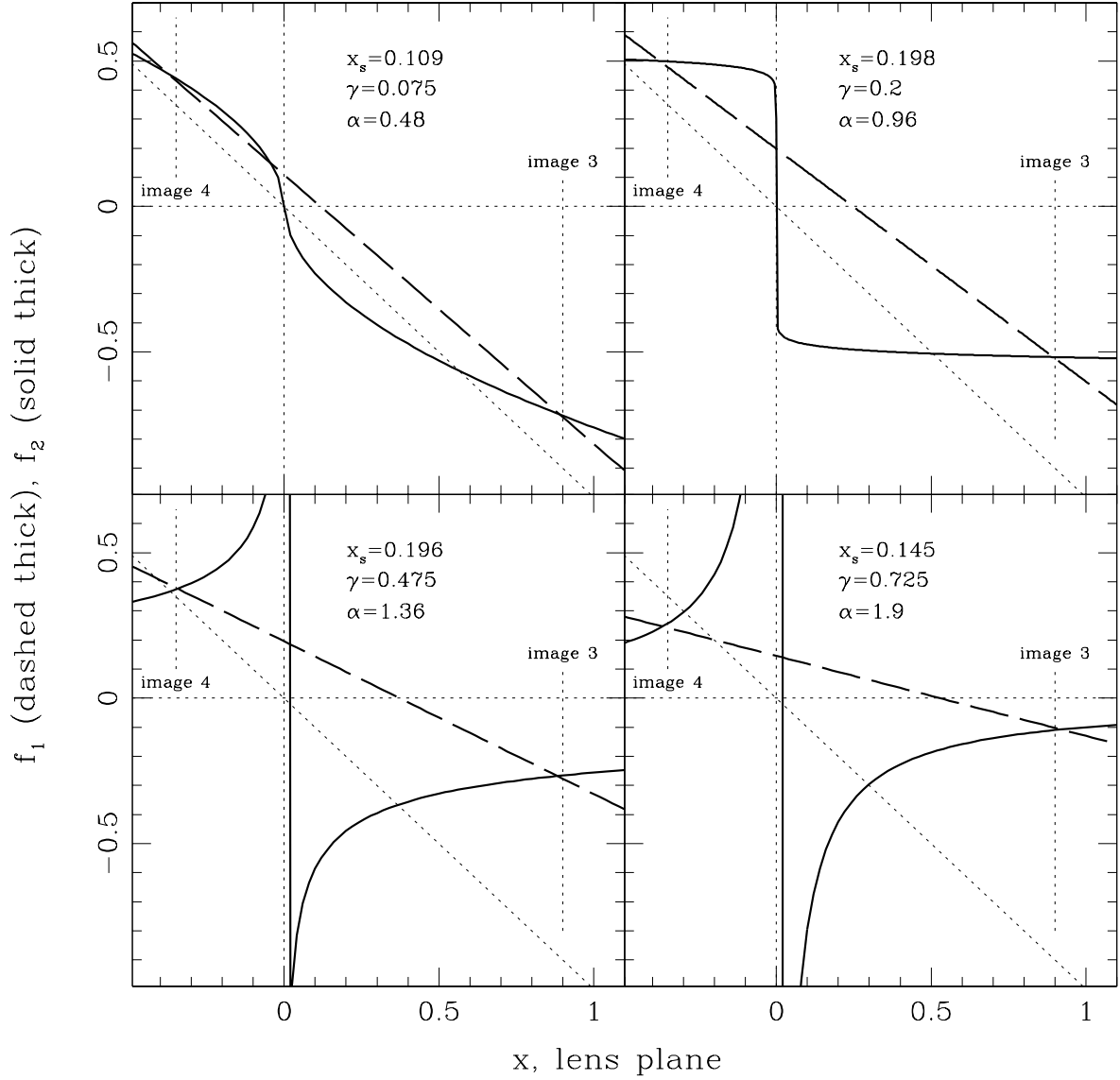}
\caption
{\footnotesize
B1422 simplified: one dimensional case showing the formation of
images 3 and 4. Thick dashed lines represent $f_1=x_I(\gamma-1)+x_S$
and thick solid lines represent $f_2=\mp A {x_I}^{1-\alpha}$; they
intersect at image locations. Four combinations of ($\gamma$, $\alpha$) 
are shown in the four panels, as labeled. The thin dotted lines show
the axes, the diagonals, and image locations.
\label{one_d}}
\end{figure}

Consider a simplified analytic model of an axisymmetric lens, with the 
QSO source, lens, group center, and the images all lying on the $x$-axis.
Let the mass distribution in the main galaxy lens around the image-ring
be given by a power law, $\rho\propto x^{-\alpha}$. 
Its contribution to the lensing potential is 
$\Phi_{\rm gal}\propto x^{2-\alpha}$, and the deflection angle is 
$d\Phi/dx=\pm A x^{1-\alpha}$, where $+$ ($-$) stands for the image formed on
the positive (negative) $x$-axis, and $A$ lumps together all the constants.
The total lensing potential is
\begin{displaymath}
\Phi={A\over(2-\alpha)}x^{2-\alpha} + {\gamma\over 2} x^2,
\end{displaymath}
where we have discarded terms that are zero on the $x$-axis.
The lens equation is
\begin{displaymath}
x_I(\gamma-1)\pm A {x_I}^{1-\alpha}+x_S=0.
\end{displaymath}
If we define $f_1=x_I(\gamma-1)+x_S$ and $f_2=\mp A {x_I}^{1-\alpha}$
and plot them against lens plane coordinate $x$, then images are
formed at the intersections. The four panels of Fig.~\ref{one_d} show 
models with four different combinations of ($\gamma$, $\alpha$):
$(0.075,0.48)$, $(0.2,0.96)$, $(0.475,1.36)$, and $(0.725,1.9)$.
All these models satisfy the locations of images 3 and 4, and the 
condition that image 4 arrives after 3. The vertical offset of $f_1$ 
is the source location, $x_S$, while external shear makes $f_1$
shallower. The difference in galacto-centric distances of images 
3 and 4 is given by $|x_3+x_4|$.

Figure~\ref{one_d} graphically demonstrates that in a system where
images are formed at very different galacto-centric distances (like 3
and 4 in B1422) there is a limit to the shallowness of the galaxy's
density profile index, $\alpha$. It is clear from the figure that some
non-zero $x_S$ is needed to produce large $|x_3+x_4|$. For shallow density
profiles (such as the one in the upper left panel) $f_2$ tends to become 
the diagonal through the II and IV quadrant. This makes it very
difficult to have $f_2$ intersect $f_1$ while keeping $|x_3+x_4|$ large.
The lower limit on $\alpha$ is around $\sim 0.5$. 

The figure also shows that $\alpha$ and $\gamma$ in B1422 are correlated:
as profile slopes get steeper, shear must increase as well.
For $\alpha>1$  (profile steeper than isothermal) $f_2$ become hyperbolic, 
and a substantial flattening of $f_1$ (i.e. large external shear) 
is required to produce large $|x_3+x_4|$. Note that large shear is
necessary for this; a large $x_S$ by itself is not sufficient. 

As long as profile is steeper than a minimum $\alpha$ value,
one-dimensional lensing case does not prefer large $\gamma$, large
$\alpha$ solutions over small $\gamma$, small $\alpha$: in
Fig.~\ref{one_d} $(\gamma,\alpha)=(0.075,0.48)$ is as good a fit as
$(0.725,1.9)$.  However, when we consider two dimensions, images 1 and
2, set an upper limit on the external shear because too
large a shear will tend to place images 1, 2 and 3 in a straight line,
perpendicular to the direction towards external shear. Since observed
images 1,2 and 3 are not in a straight line, the shear is constrained
from above. Furthermore, if the tensor magnification information is
included in modeling (as is done in Section~\ref{pixelated}) the relative
orientation of these for images 1 and 2 also reduced the amplitude of 
possible shear. 

We can intuitively understand the effect of external shear on the time
delays $\Delta t$ as follows.  If the shear source is far from the
image region, the shear can be represented by constant external shear,
which provides linear deflection angles at the images. It corresponds
to tilting the side of the arrival-time surface closest to the source
of the shear upwards, which moves the zero points in $\Delta t$ away
from the shear source.  Images lying on the line through the shear
source, QSO source and the main galaxy are most affected by this tilting
(images 3 and 4 in B1422), while those lying perpendicular to that
line are least affected (1 and 2).  Thus images 3 and 4 give a well
constrained $\Delta t$, precisely because they are most sensitive to
external shear.

To sum up, the fortuitous geometry of B1422 and large external shear
imply a relation between profile slope and shear magnitude, and impose
a limit on the shallowness of the profile slope of $\alpha\sim 0.5$, or
a lower limit on shear of $\gamma\sim 0.1$, consistent with the shear
estimates based on the physical properties of the group 
(Section~\ref{ext_shear}). In addition, images 
1 and 2 impose a limit on the steepness of the profile slope 
(through an upper limit on shear). As a result, $\Delta t_{34}$
should be well constrained.

\section{Pixelated models}\label{pixelated}

We now consider more detailed models of galaxy mass distribution in
the lens plane.

Given the difficulties caused to previous models of B1422 
by the observed flux
ratios, and especially since we now need to fit tensor magnifications,
it is clear that more general models than those are necessary.
Moreover, to properly estimate uncertainties, it is necessary to
explore not just a few models but large ensembles of
them.\footnote{``Aggressively explore all other classes of models'' in
the words of \citet{bk96}.}  The pixelated lens reconstruction method,
as developed in \citet{sw97} and \citet{ws00}, was designed for this
purpose.  The idea is to model the lens as a sum of mass tiles or
pixels, with two kinds of constraints: the ``primary constraints'' are
that the lensing data should be fitted exactly; ``secondary
constraints'' require the mass distribution to be centrally
concentrated, not excessively elongated, and optionally
inversion-symmetric.  Here we follow the method of generating
ensembles of models as detailed in the latter paper, but with two
minor modifications necessitated by the type of data.

The first modification is needed because the time delay measurements
are currently very uncertain.  So instead of including them, we run
the reconstruction code with one time delay ($\Delta t_{34}$, say)
set to a fiducial value $d_{\rm fid}$.  The code then generates an
ensemble of models, each model having its own $h$ and its own set of
$\Delta t_{ij}$.  Because of the scaling properties of the lensing
problem, the only dependence of the results on $d_{\rm fid}$ will be
that all the model $h^{-1}$ and $\Delta t_{ij}$ will be proportional
to it.  Hence $h\Delta t_{ij}$ will be independent of $d_{\rm fid}$.

A second modification is needed to incorporate tensor magnifications.
To do this, we first reconstruct the magnification matrix {\bf M} at
each image position from the data in Table~\ref{tbl}.  We have
\begin{equation}
{\bf M} = {\bf R}(-\omega) \pmatrix{\sqrt{Mr}&0\cr0&\sqrt{M/r}}
          {\bf R}(\omega)
\label{tensormag}
\end{equation}
where $M$ is the scalar magnification (proportional to the flux), $r$
is the axis ratio (if we assume the source is circular), $\omega$ is
the orientation ($\rm PA+90^\circ$), and {\bf R} is a rotation matrix.
The signs of the square roots in equation~\ref{tensormag} depend on
image parity: for minima both roots will be positive; for saddle
points one will be negative.  Having reconstructed {\bf M} at the
image positions, we write the observed upper and lower bounds on the
axis ratio as bounds on the ratio of the diagonal elements of {\bf M}
(see \citet{asw98} for an application of this technique to arclets in
cluster lensing).  Having constrained the axis ratio, we set the flux
ratio by introducing a fictitious quad slightly displaced from the
actual one; in particular, fictitious images displaced by
\begin{equation}
\pmatrix{\Delta\theta_x\cr\Delta\theta_y} = {\bf M} \pmatrix{\epsilon\cr0}
\label{exim}
\end{equation}
correspond to a fictitious source displaced by $\epsilon$ along the
$x$-axis of the source plane.  The code is not told $\epsilon$ itself,
just the displacements it maps to; in this way the magnification
ratios are constrained, but not the absolute magnifications.

We have generated three ensembles of models.  Each ensemble contains
100 models sampling the allowed `model space' given a certain set of
constraints.

\begin{figure}[p!]
\epsscale{0.4}
\plotone{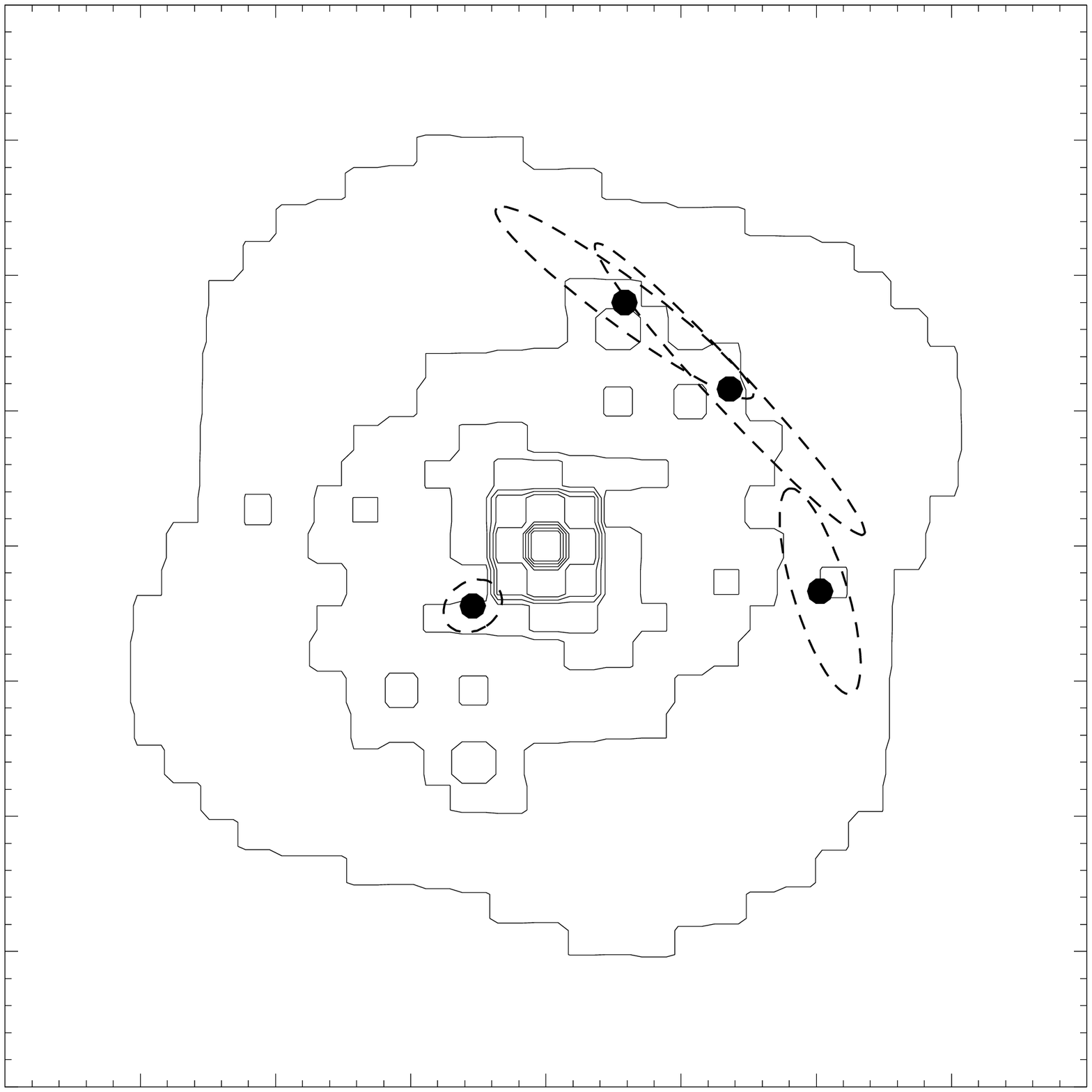}
\caption
{\footnotesize
Ensemble average of 100 mass maps including tensor magnification and
assuming inversion symmetry. The pixel size is evident. Contours are
in steps of $1\over 3$ of the critical density for lensing, 
with the outermost contour at $1\over 3$ of the critical density.  
The filled circles
indicate the image, and dashed ellipses around them indicate the
tensor magnification.  But note that we have made the areas of the
ellipses proportional to $\sqrt{M}$ rather than $M$.
\label{mass_s}}

\epsscale{0.48}
\plotone{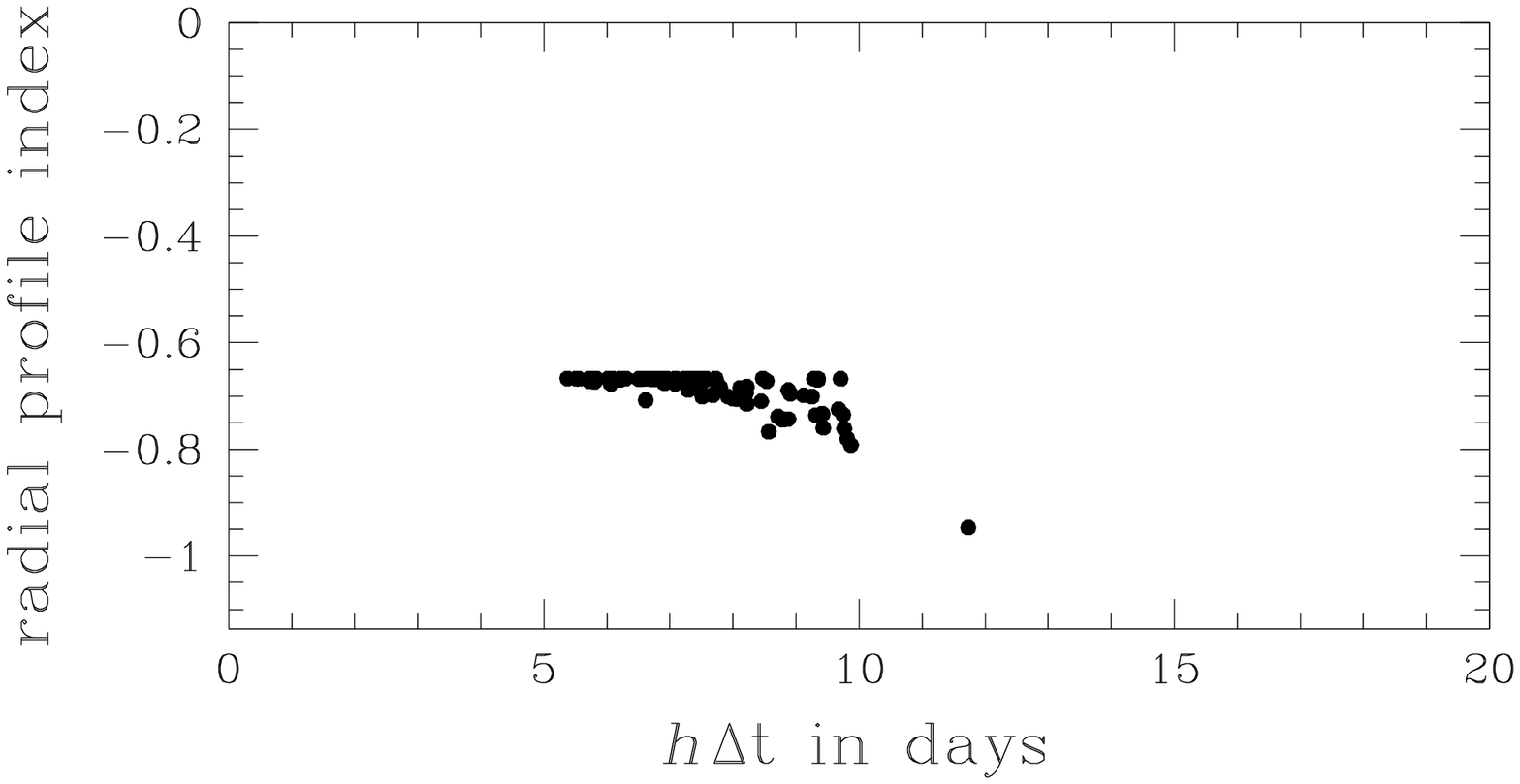}
\plotone{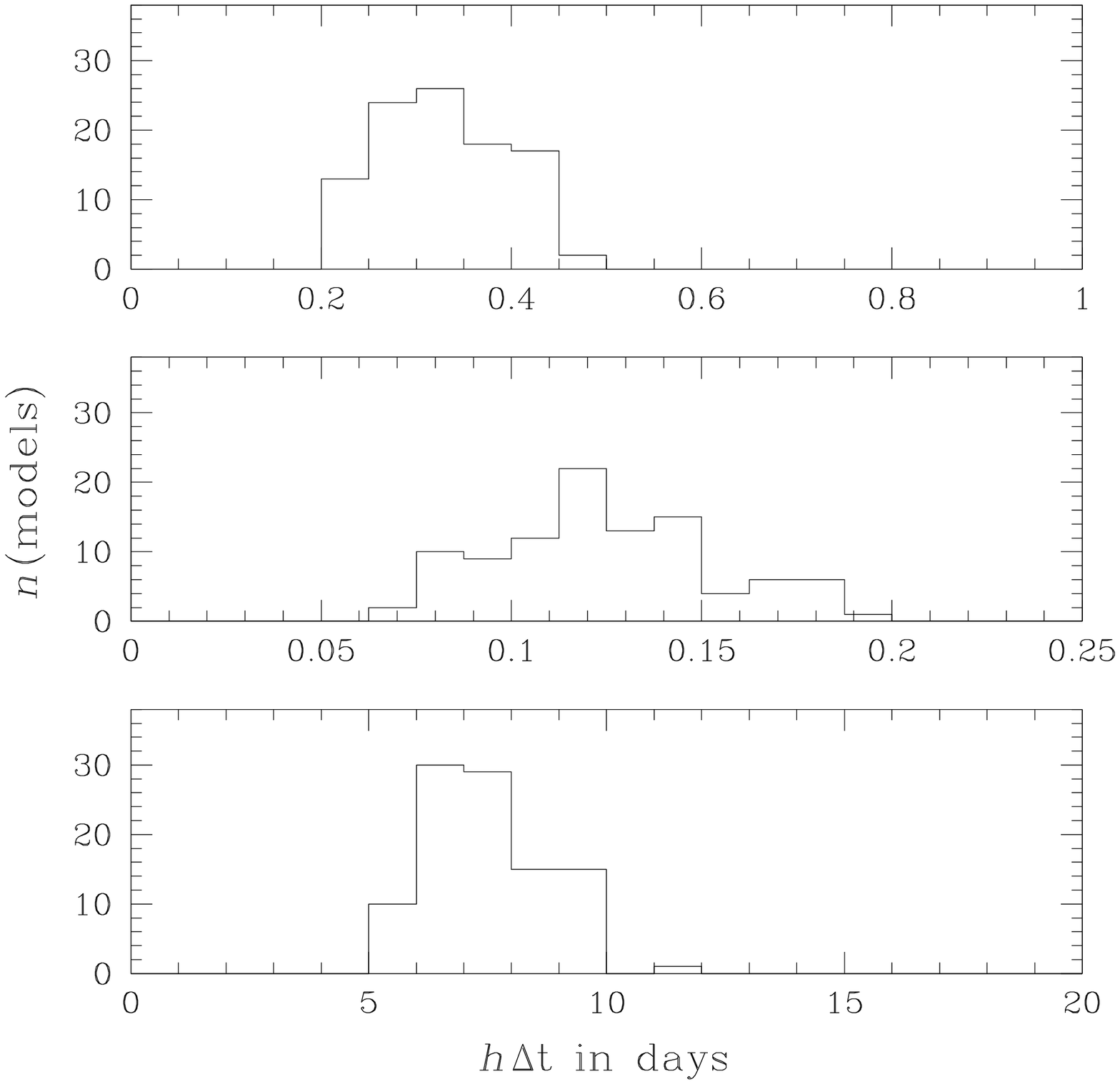}
\caption
{\footnotesize
{\bf Left panel:} Predicted time delays (in the form $h\Delta t$ in
days) between images 3 and 4, for the same ensemble as in
Fig.~\ref{mass_s}.  {\bf Other panels:}~Predicted time delays between
different images (going downwards, 1--2, 2--3, and 3--4) for the same
ensemble. Note that the horizontal scale varies between panels.
\label{delay_s}}
\end{figure}

The first ensemble uses the data from Table~\ref{tbl}, together with
the generic secondary constraints \citep{ws00} including inversion
symmetry of galaxy-lens. Figure~\ref{mass_s} shows the
ensemble-average mass map, while Fig.~\ref{delay_s} shows the
distribution of predicted time delays.  (The `radial profile index' in
Fig.~\ref{delay_s} corresponds approximately to $-\alpha$ in the
previous section.)  The qualitative results are just as expected
from our preceding discussion: (i) images 1,2,3 are very close in
$\Delta t$, with 4 arriving much later; (ii) $\alpha$ is nearly
confined to a narrow range 0.6--0.8; (iii) while $\Delta t_{34}$ has
a broad peak with median $7.3h$ days, it is much better constrained
than in models of other well-studied quads \citep{ws00}.

\begin{figure}[p!]
\epsscale{0.4}
\plotone{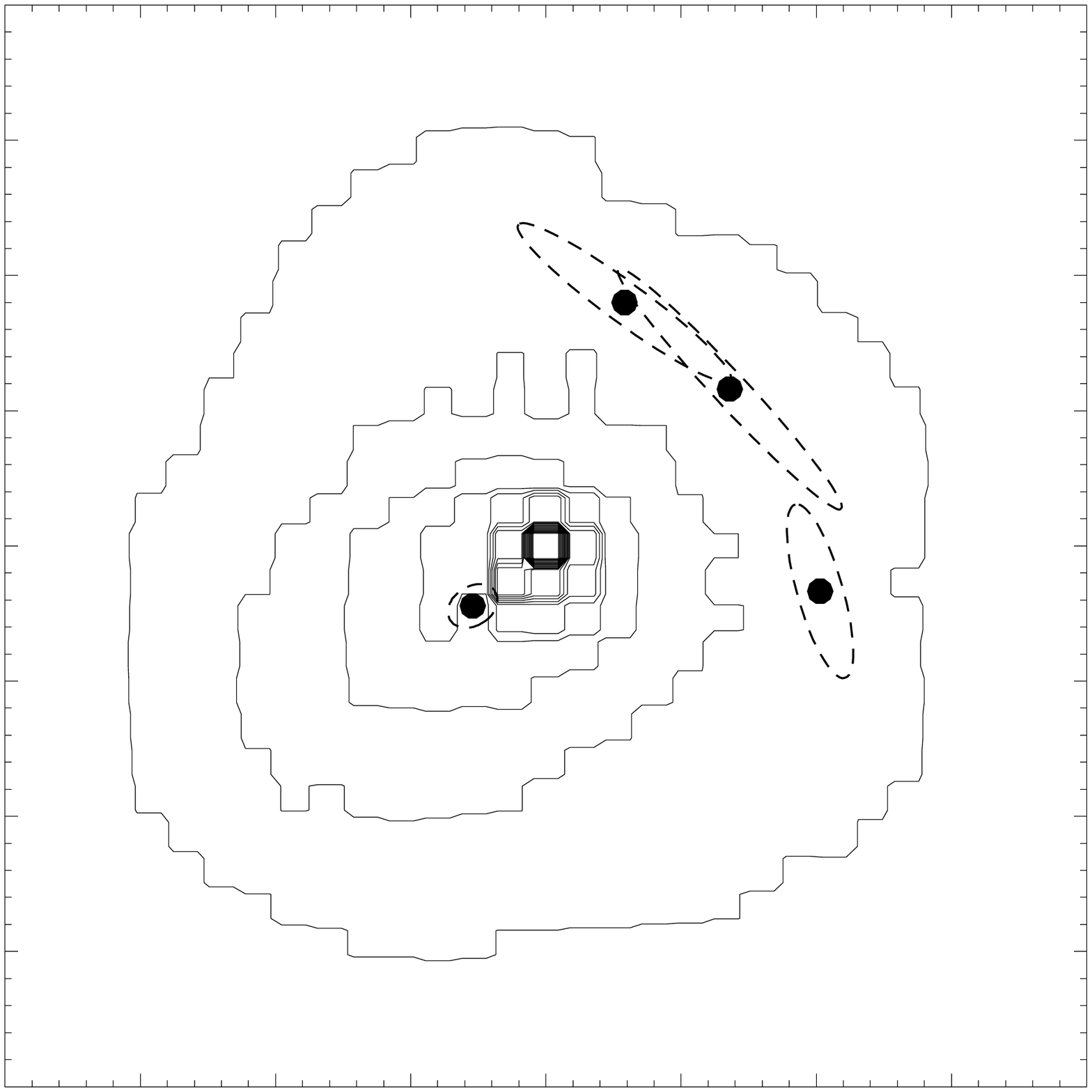}
\caption
{\footnotesize
Like Fig.~\ref{mass_s}, but without the inversion-symmetry constraint.
\label{mass_a}}

\epsscale{0.48}
\plotone{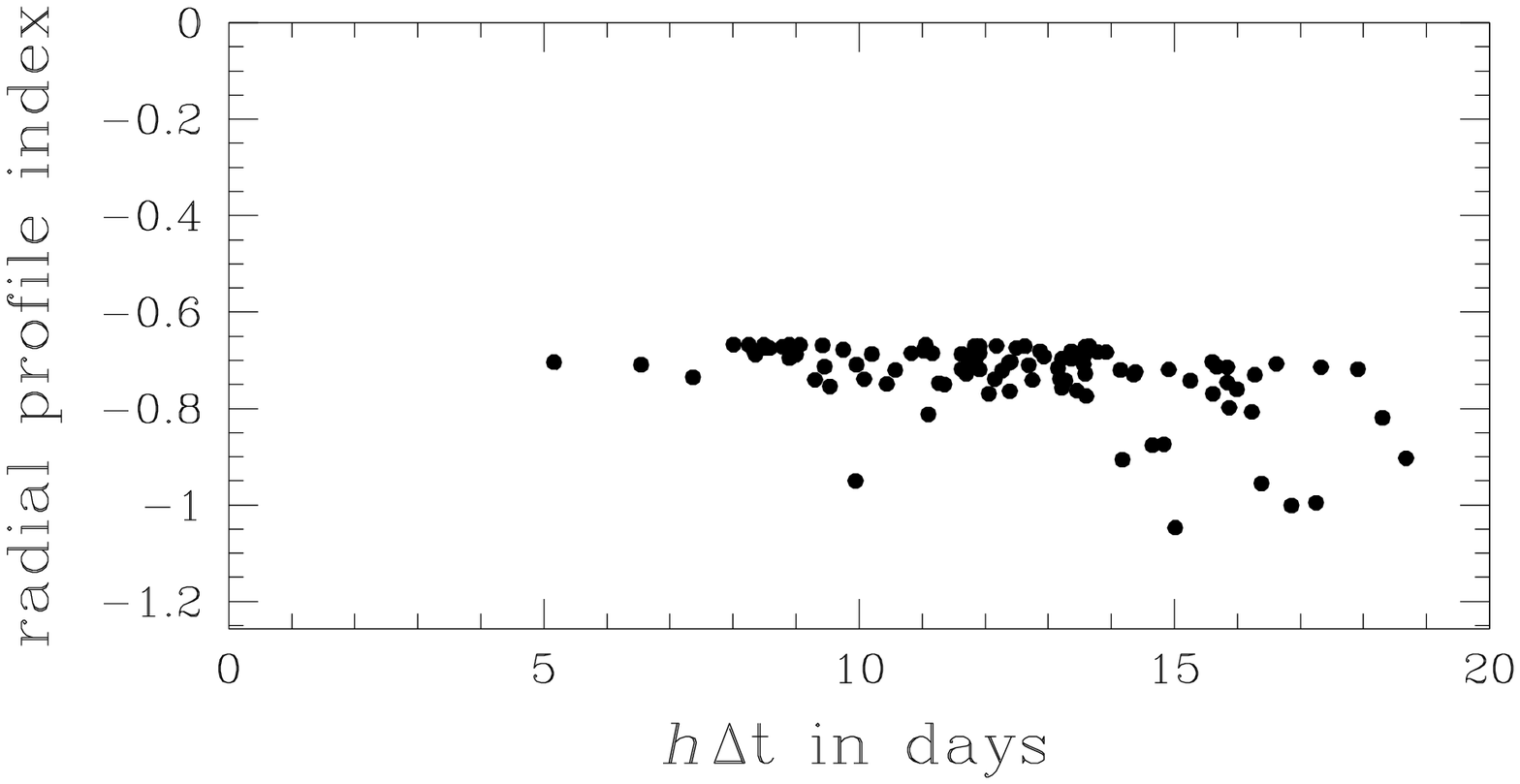}
\plotone{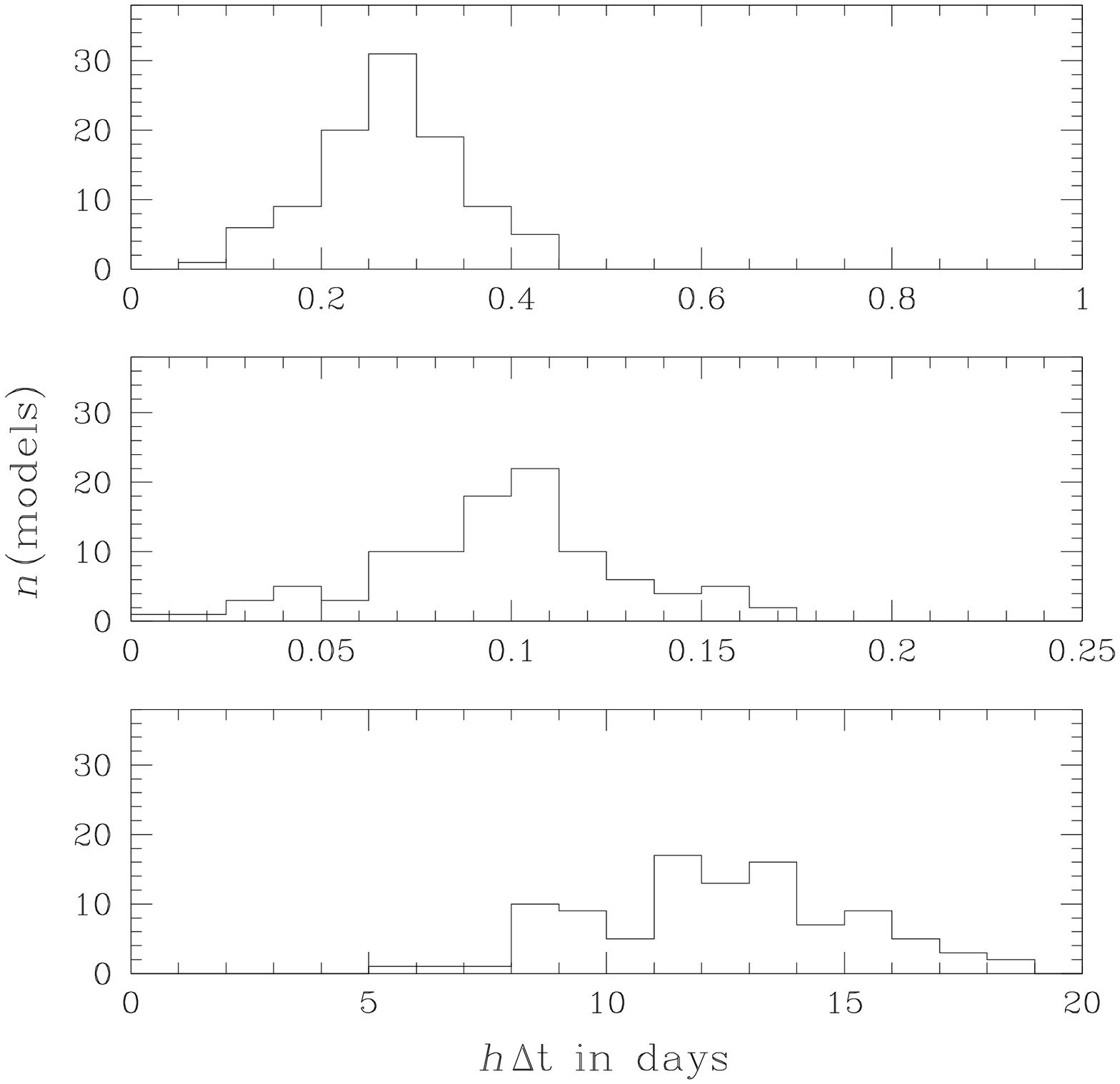}
\caption
{\footnotesize
Like Fig.~\ref{delay_s}, but for models without the inversion
symmetry constraint.
\label{delay_a}}
\end{figure}

For the second ensemble we dropped the inversion-symmetry constraint.
Figures~\ref{mass_a} and \ref{delay_a} show the results.
Qualitatively, the results are similar, but the predicted $\Delta
t_{34}$ shifts to a higher range (median $12.4h$ days).  One must keep
in mind that observed image properties tightly constrain the position
angle of the shear, not its radial location.  So
non-inversion-symmetric modeling of lens systems with large external
shear will produce mass profiles that are artificially elongated
towards the source of external shear.  This is seen in
Fig.~\ref{mass_a}. The elongation results because the model cannot
decide how far away to place the source of the shear.  Thus, for
systems with non-disturbed galaxy lenses and large external shear the
inversion-symmetric models are probably somewhat more trustworthy than
non-inversion-symmetric ones.

\begin{figure}[tbp!]
\epsscale{0.48}
\plotone{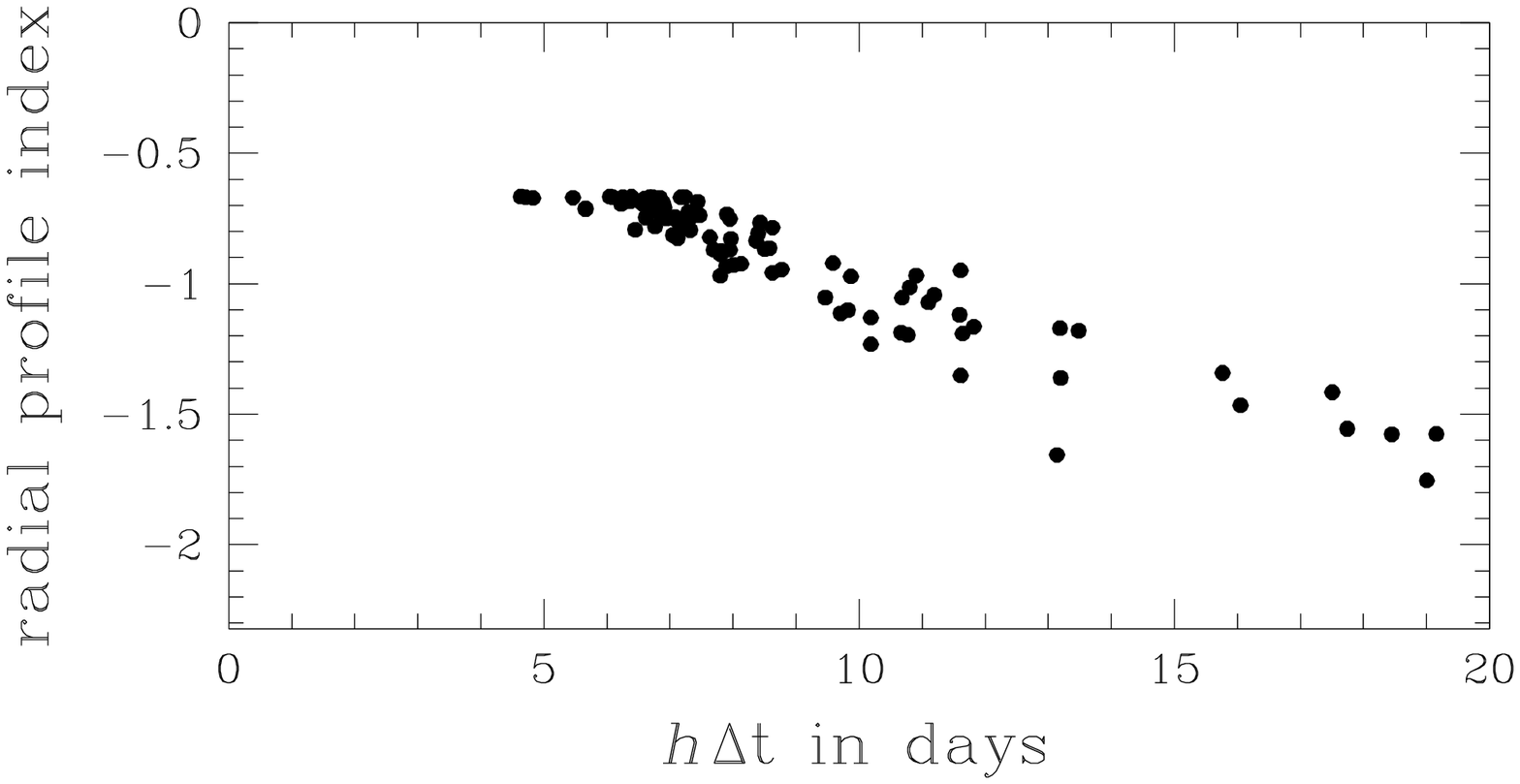}
\plotone{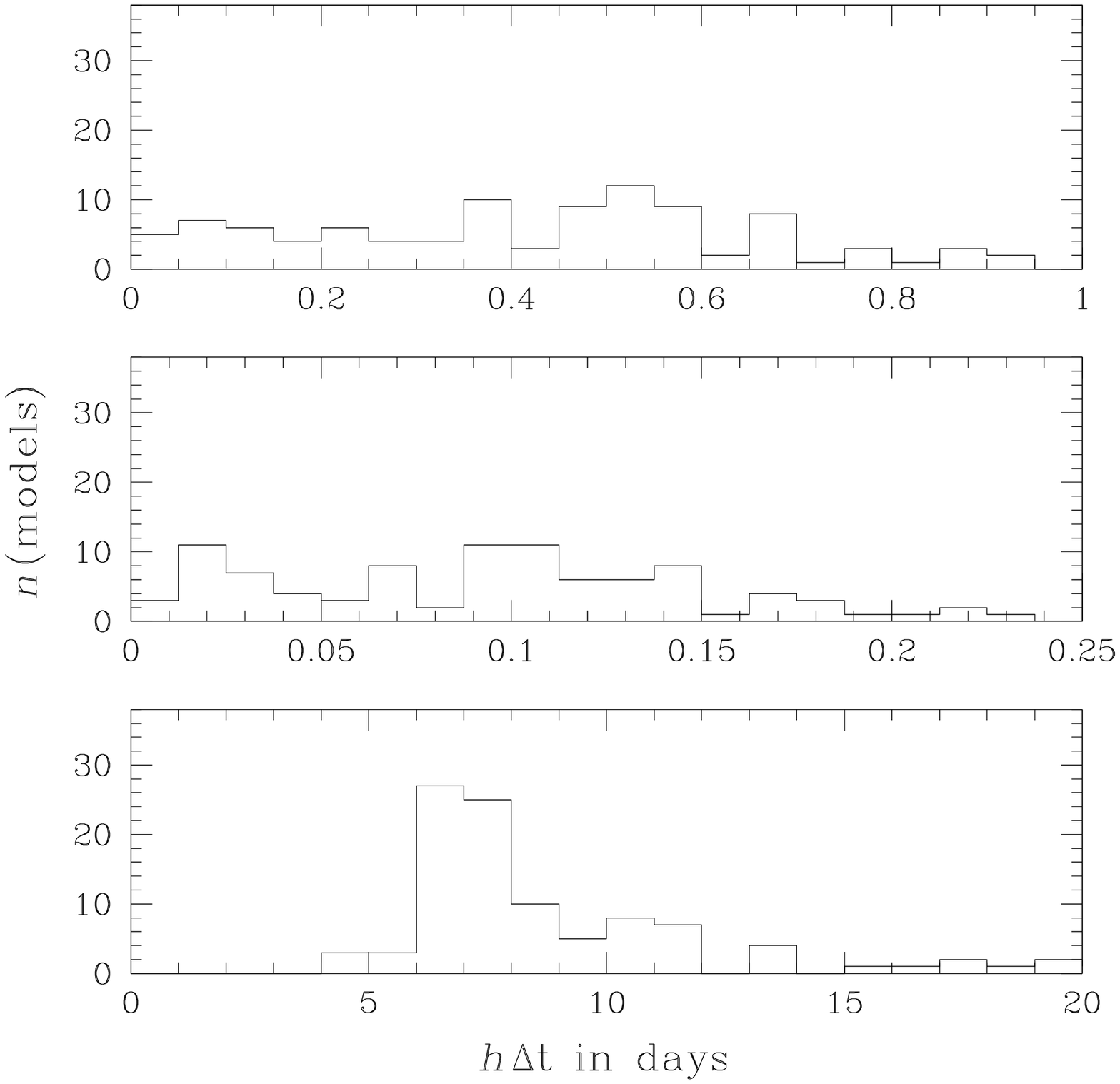}
\caption
{\footnotesize
Like Fig.~\ref{delay_s}, but for models without magnification
constraints.
\label{delay_n}}
\bigskip\hrule
\end{figure}

For the third ensemble we put back the inversion-symmetry constraint,
but drop the magnification constraint.  Figure~\ref{delay_n} shows the
time-delay predictions.  (The ensemble-average mass map looks similar
to Fig.~\ref{mass_s} and we have not included it here.)  Comparison of
Fig.~\ref{delay_n} and Fig.~\ref{delay_s} shows what magnification
information adds to modeling.  Basically, magnification measurements
put constraints on the mass distribution in the neighborhood of the
images.  Time delays between nearby images becomes better
constrained. Thus Fig.~\ref{delay_n} has a much larger spread in
$\Delta t_{12}$ and $\Delta t_{23}$ than Fig.~\ref{delay_s}. Without
magnifications, some models' $\Delta t_{12}$ or $\Delta t_{23}$ can
become zero: magnifications for the images in question can get
arbitrarily large, effectively merging those images.  Also, the upper
bound on $\alpha$ that we argue comes from images 1 and 2 is greatly
weakened.  On the other hand, time delays between distant images is
hardly affected: as anticipated in Section~\ref{simple}, $\Delta
t_{34}$ in Fig.~\ref{delay_n} has a median ($7.8h$ days) and range
very similar to those in Fig.~\ref{delay_s}.

\section{Conclusions}

In the Introduction, we mentioned four unusual features that make
B1422 a particularly interesting lens.  We now summarize what the
results of this paper indicate about each of these features.

First, we have the surprising result that being dominated by external
shear from group galaxies makes the lens {\em better\/}
constrained. The predicted time delays, in particular the longest
delay $\Delta t_{34}$, though it has a broad range, is narrower than
in other comparable systems. The fortuitous alignment of the source
displacement and the shear appears to help.

Second, one detail of the group contribution is important.  Of two
situations: (i)~the lensing galaxy has $180^\circ$-rotation symmetry
and the shear comes from relatively distant group members, and
(ii)~the lensing galaxy is asymmetric and elongated along the group
direction, the second case gives $\Delta t_{34}$ 50\% longer. Optical
and X-ray (see Fig.~\ref{fig:xray}) images of the galaxy and group
are not conclusive on this point.

Third, tensor magnifications from fluxes and VLBI constrain time
delays between nearby images; but remarkably they have no discernible
effect on longer time delays, $\Delta t_{34}$. 
The physical reason is not hard to
appreciate: magnification is essentially the second derivative of the
time delay, hence time delays between widely separated images tend to
wash out the sort of local variations of density that cause
differences in magnifications.

Fourth, comparison with time-delay measurements is still problematic
at present.  \citet{pat01} report $\Delta t_{12}=7.6\pm2.5\rm d$,
whereas our reconstructions predict $\Delta t_{12}\sim0.4h\rm d$. The
difficulty is that our predicted values would not be expected to show
up in the Patnaik \& Narasimha data, which sample at $\sim 4\rm d$;
instead, aliases of our predicted values would show up.  Even with
closely-sampled data, a $\Delta t_{12}\sim0.4h\rm d$ is unlikely to
be measurable because it would need intrinsic brightness variations on
the scale of hours.  On the other hand, the predicted $\Delta
t_{34}\sim 5-10\rm d$ is a convenient length for measurements; but
unfortunately it involves image 4, whose flux is $\sim30$ times
smaller.

In summary, we conclude that B1422 is a very interesting lens, but
probably not the sought-after golden lens.

\acknowledgements
We thank the CASTLES team \citep{castles}
for supplying us with the near-IR HST/NICMOS image used in 
Fig.~\ref{fig:xray}.

\end{document}